\documentclass[manuscript,aps,epsfig,graphicx]{revtex4}
\usepackage[dvips]{graphicx}
\newcommand{\be}{\begin{equation}}
\newcommand{\ee}{\end{equation}}

\newcommand{\bea}{\begin{eqnarray}}
\newcommand{\eea}{\end{eqnarray}}
\newcommand{\bd}{\begin{displaymath}}
\newcommand{\ed}{\end{displaymath}}
\newcommand{\bi}{\begin{itemize}}
\newcommand{\ei}{\end{itemize}}
\newcommand{\bc}{\begin{center}}
\newcommand{\ec}{\end{center}}
\newcommand{\bfl}{\begin{flushleft}}
\newcommand{\efl}{\end{flushleft}}
\newcommand{\bfr}{\begin{flushright}}
\newcommand{\efr}{\end{flushright}}
\newcommand{\f}{\frac}

\def\bk{{\vec{k}}} \def\bq{{\bf q}}

\def\ra{\rightarrow}
\def\6{\partial} \def\a{\alpha} \def\b{\beta}
  \def\ve{\varepsilon} 
  
  \def\l{\lambda}
   
 \def\ss{\sigma} 

\def\o{\omega}  \def\D{\Delta}
 \def\L{\Lambda} 
  \def\O{\Omega}

\def\={\!\!\!&=&\!\!\!}
\def\+{\!\!\!&&\!\!\!+~}
\def\-{\!\!\!&&\!\!\!-~}

\begin{document}
\author{I. \c{T}ifrea}
\affiliation{Department of Theoretical Physics, "Babes-Bolyai"
University, 3400 Cluj, Romania}
\affiliation{Department of Physics
and Astronomy, University of Iowa, Iowa City, IA 52242, USA}
\author{J.A. Budagosky Marcilla}
\author{J.J. Rodr\'{\i}guez--N\'u\~nez}
\affiliation{Departamento de F\'{\i}sica--FACYT, Universidad de
Carabobo, Valencia, Edo. Carabobo, Venezuela}
\date{\today}
\title{Ginzburg-Landau Expansion in Non-Fermi Liquid
Superconductors: Effect of the Mass Renormalization Factor}
\begin{abstract}
We reconsider the Ginzburg-Landau expansion for the case of a
non-Fermi liquid superconductor. We obtain analytical results for
the Ginzburg-Landau functional in the critical region around the
superconducting phase transition, $T\leq T_c$, in two special
limits of the model, i.e., the spin-charge separation case and the
anomalous Fermi liquid case. For both cases, in the presence of a
mass renormalization factor, we derived the form and the specific
dependence of the coherence length, penetration depth, specific
heat jump at the critical point, and the magnetic upper critical
field. For both limits the obtained results reduce to the usual
BCS results for a two dimensional s-wave superconductor. We
compare our results with recent and relevant theoretical work. The
results for a d--wave symmetry order parameter do not change
qualitatively the results presented in this paper. Only numerical
factors appear additionally in our expressions.
\end{abstract}
\maketitle
\section{Introduction}

\indent The discovery of high temperature superconductivity (HTSC)
in 1986 by Bednorz and M\"{u}ller\cite{BM} has caused a lot of
enthusiasm among the physics community. Today, after 15 years from
their discovery, these materials are still not completely
understood from the theoretical point of view. Being a part of the
largest family of strongly correlated electron systems, the
cuprates present anomalous properties both in the normal and
superconducting phase. As a consequence, standard theories, the
Landau theory of the Fermi liquid and the BCS theory of the
superconducting state, fail to correctly describe the physical
properties of these materials. As alternatives, in the case of the
normal state, several phenomenological models have been proposed
in order to explain their nonmetallic behavior. \cite{theory}
Despite the fact that the superconducting transition occurs at
high temperatures, a characteristic of the ordered phase is the
presence of the electron pairs, leading to the idea that a
modified BCS theory is appropriate for the description of their
superconducting state. We mention that Franz and
Tesanovic\cite{TesaQED3} has put forward a phase fluctuation model
for the pseudogap state of cuprate superconductors, which includes
a d--wave order symmetry, generating non--Fermi liquid behavior
(see their Fig. 1).

\indent In the following, for the description of the normal state
we will adopt the model proposed by Anderson\cite{PWA,Anderson}
which is based on the hypothesis that a two dimensional (2D)
system can be described by a Luttinger liquid type theory, similar
to the one dimensional (1D) case. The generalization of the
Luttinger liquid for the 2D case involves the use of the following
Green's function in order to describe the free particles in the
normal state of cuprates:
\begin {equation}\label{nflG}
G(\vec{k},i\omega_n)=\frac{g(\alpha)
e^{-i\pi\alpha/2}}{\omega_c^\alpha (i\omega_n - \b \eta
\varepsilon_{\vec{k}})^{1/2} (i\omega_n - \b
\varepsilon_{\vec{k}})^{1/2-\alpha}}\;,
\end {equation}
where $\o_c$ is a cut-off frequency, $\eta=u_\ss/u_\rho<1$ is the
ratio of the spin and charge velocities in the system, $\a$ is the
non universal exponent related to the anomalous Fermi surface,
$\b=2/(\eta+1-2\alpha)$ is the mass renormalization factor, and
$g(\a)=\pi\a/[2\sin(\pi\a/2)]$. Relations between the different
parameters entering Eq. (\ref{nflG}) can be obtained by studying
different general properties of the Green's function. For example
the form of the function $g(\a)$ was obtained by the use of the
first sum rule.\cite{yin,massren} Based on the formalism proposed
by Nolting\cite{Nolting}, the necessity of the mass
renormalization factor was predicted in Ref. \cite{massren} using
high order sum rules.

\indent Different choices of the given parameters, $\a$ and
$\eta$, give us the possibility to distinguish between several
different regimes incorporated in the Green's function (Eq.
(\ref{nflG})). The most general case is the one of an anomalous
spin-charge separation Fermi liquid characterized by $\eta\neq 1$
and $\a\neq 0$. Unfortunately, for this general case, analytical
results are very difficult to obtain. The spin--charge separation
liquid is obtained when $\eta\neq 1$ and $\a=0$, when the Green's
function is very similar with the one characterizing the Luttinger
liquid. The case of an anomalous Fermi liquid is obtained for
$\eta=1$ and $\a\neq 0$. In this case, the spectral function
$A(\bk, \o)=-Im G(\bk, \o)/\pi$ satisfies the homogeneity relation
$A(\L\bk, \L\o)=\L^{-1+\a}A(\bk, \o)$ with an exponent $\a>0$. As
a general feature the usual Fermi liquid limit is obtained by
simply considering $\eta=1$ and $\a=0$.

\indent Previously, this model was used in order to investigate
the superconducting state in non-Fermi liquid
systems.\cite{IonelPh,nelu,muthu,sudbo} Several properties of the
superconducting state were investigated assuming the validity of
the Gorkov's equations, with the usual normal state Fermi liquid
Green's function replaced by the non-Fermi liquid one given by Eq.
(\ref{nflG}). For the case of the anomalous Fermi liquid
($\a\neq0$, $\eta=1$) there have been previous analysis of the
Ginzburg-Landau parameters in the framework of fluctuation
theory\cite{flnonfermi} and general Ginzburg-Landau functional
theory\cite{Moca}, but these two approaches missed the role of the
mass renormalization factor, leading to uncomplete results.

\indent In this paper, based on the Ginzburg--Landau functional
formalism, we investigate both the spin--charge separation liquid
and the anomalous Fermi liquid in the superconducting state and we
evaluate the coherence length, the penetration depth, the specific
heat jump at the critical point, and the magnetic upper critical
field.

\section{General formalism of the Ginzburg-Landau functional}

Let us first consider the general case of a $s$--wave
superconductor. In the standard Ginzburg-Landau expansion, the
difference between the superconducting and normal state free
energy can be written as:
\be\label{freeen}
F_S-F_N=A|\D_\bq|^2+q^2C|\D_\bq|^2+\f{B}{2}|\D_\bq|^4\;,
\ee
where $S$ denotes the superconducting state, $N$ the normal state,
$\D_\bq$ is the Fourier transform of the order parameter, and A,
B, C are the Ginzburg-Landau coefficients. Following Ref.
\cite{sadovskii} we can express these coefficients as:
\bea\label{Acoef}
A \equiv \frac{1}{V} - T\sum_n\int \frac{d^2k}{(2\pi)^2} \left[
{\cal G}(\vec{k},i\omega_n) {\cal G}(-\vec{k},-i\omega_n)\right.\nonumber\\
+ \left. {\cal F}(\vec{k},i\omega_n) {\cal F}(-\vec{k},-i\omega_n)
\right]\;,
\eea
\bea\label{Bcoef}
B \equiv T_c \sum_n\int \frac{d^2k}{(2\pi)^2} \left[
{\cal G}(\vec{k},i\omega_n) {\cal G}(-\vec{k},-i\omega_n)\right.\nonumber\\
+\left. {\cal F}(\vec{k},i\omega_n) {\cal F}(-\vec{k},-i\omega_n)
\right]^2\;,
\eea
and $C$ as the coefficient of the $q^2|\D_\bq|^2$ term in the
Taylor expansion of the following term:
\bea\label{Ccoef}
- T_c \sum_n\int \frac{d^2k}{(2\pi)^2} \left[ {\cal
G}(\vec{k}+\frac{\vec{q}}{2},i\omega_n)
{\cal G}(-\vec{k}+\frac{\vec{q}}{2},-i\omega_n)\right. \nonumber\\
+\left. {\cal F}(\vec{k}+\frac{\vec{q}}{2},i\omega_n) {\cal
F}(-\vec{k}+\frac{\vec{q}}{2},-i\omega_n) \right]\;.
\eea
\indent For the case of a $s$--wave superconductor ${\cal G}(\bk,
i\o_n)$ and ${\cal F}(\bk, i\o_n)$ represent the normal and
anomalous Green's functions, respectively, in the superconducting
state.\cite{gorkov} In Eq. (\ref{Acoef}), $V$ represents the
absolute value of the attractive interaction leading to the
superconducting phase transition. In general, in order to evaluate
$A$, the interaction potential is replaced using the critical
temperature equation, leading for  the $s$--wave case to the
following value:
\be\label{valA}
A_0=N_0\f{T-T_{c0}}{T_{c0}}\;,
\ee
where $T_{c0}$ stays for the standard BCS critical temperature and
$N_0$ the free electron gas density of states at the Fermi level.
The other two coefficients, $B$ and $C$, can be calculated after
some simple but tedious algebra. The results are:
\be\label{valB}
B_0=\f{7\zeta(3)N_0}{8\pi^2T_{c0}^2}\;,
\ee
and
\be \label{valC}
C_0=\f{7\zeta(3)v_F^2N_0I(\theta)}{16\pi^2T_{c0}^2}\,
\ee
where $\zeta(z)$ is the Riemann function, $v_F$ is the Fermi
velocity and
\be\label{I}
I(\theta)=\left\{ \begin{array}{ll}\frac{1}{2}, & \mbox{2D case}\\ \\
\frac{1}{3}, & \mbox{3D case}\end{array}\right.\;.
\ee
\indent Of course for the correct dimensional evaluation of the
Ginzburg-Landau coefficients, the density of states $N_0$ should
be considered according to the dimensionality. As we expected
$A=0$ gives the mean field critical temperature of the
superconducting phase transition, whereas $B$ and $C$ are weakly
temperature dependent and are evaluated at $T=T_{c0}$.

Standard BCS theory considered a simple $s$-wave symmetry of the
order parameter in standard superconductors. However, at the
present time is generally accepted that in HTSC the order
parameter symmetry is more of $d$-wave type.\cite{tsuei}
Accordingly, an evaluation of the Ginzburg-Landau coefficients for
HTSC should consider a momenta dependent order parameter
$\D(\bk)=\D\psi(\bk)$, where $\psi(\bk)$ is a coefficient which
includes the symmetry factor. For $s$-wave superconductors this is
simple $\psi(\bk)=1$. For the 2D $d$-wave symmetry case,
$\psi(\bk)=\cos{(2\theta_\bk)}$, where
$\theta_\bk=\arctan{(k_y/k_x)}$. A calculation of the
Ginzburg-Landau coefficients in the framework of $d$-wave symmetry
is straightforward leading to the following results
\be\label{adwave}
A_0^{(d)}=K_A\;A_0,
\ee
\be\label{bdwave}
B_0^{(d)}=K_B\;B_0,
\ee
\be\label{cdwave}
C_0^{(d)}=K_C\;C_0,
\ee
where $K_A=K_C=1/2$ and $K_B=3/8$.

\indent As it is well known, the knowledge of the Ginzburg-Landau
coefficients give direct insight on several of the superconducting
state properties. Among these properties, we notice two
characteristic lengths, namely the coherence length and the
penetration depth of the magnetic field. In the following, we will
present the standard BCS results for a 2D $s$--wave
superconductor. The coherence length at a specific temperature,
$\xi_{BCS}(T)$, characterizes the size of the Cooper pair and can
be expressed as:
\be\label{xibcs}
\xi_{BCS}(T)=\sqrt{-\f{C_0}{A_0}}=0.74\f{\xi_{BCS}(0)}{\sqrt{1-T/T_{c0}}}\;,
\ee
where $\xi_{BCS}(0)=0.18v_F/T_{c0}$ represents the coherence
length at $T=0$. The other characteristic length is the London
penetration depth of a magnetic field in the superconductor, and
for the 2D s-wave BCS case is expressed as:
\be \label{lambdabcs}
\l_{BCS}(T)=\sqrt{-\f{c^2}{32\pi e^2
}\f{B_0}{A_0C_0}}=\f{1}{\sqrt{2}}\f{\l_{BCS}(0)}{\sqrt{1-T/T_{c0}}}\;,
\ee
with $\l^2_{BCS}(0)=mc^2/(4\pi ne^2)$ being the London penetration
depth at $T=0$ ($c$ stays for the usual value of the light speed
in vacuum, $n$ is the density, and $e$ represents the electron
charge).

Other two important quantities which emerge directly from the
Ginzburg-Landau coefficients are the specific heat jump at the
critical point and the upper magnetic critical field $H_{c2}$. In
terms of these coefficients, the specific heat jump at the
critical point can be expressed as:
\be\label{heatbcs}
\f{\D
C^{BCS}_v}{\O}=\f{T_{c0}}{B_0}\left(\f{A_0}{T-T_{c0}}\right)^2\;,
\ee
with $\O$ being the sample volume.  Based on this equation at the
critical point one obtain:
\be\label{spectcbcs}
\left(\f{\D C^{BCS}_v}{\O}\right)_{T_{c0}}=N_0\f{8\pi^2
T_{c0}}{7\zeta(3)}\;.
\ee
\indent The upper magnetic critical field is given by:
\be\label{magbcs}
H^{BCS}_{c2}=-\f{\phi_0}{2\pi}\f{A_0}{C_0}\;,
\ee
$\phi_0=\pi c/e$ being the quantum of the magnetic flux. The slope
of the curve for the upper critical field can be easily obtained
from Eq. (\ref{magbcs}) as:
\be\label{slopebcs}
\left|\f{dH^{BCS}_{c2}}{dT}\right|_{T_{c0}}=\f{16\pi\phi_0}{7\zeta(3)v^2_F}T_{c0}\;.
\ee
\indent Beside these important parameters for the superconducting
state, based on the superconducting order parameter fluctuations,
some important features of the normal state were investigated.
Among different properties, the presence of electron pairs in the
critical region above $T_c$ is responsible for the presence of a
gap in the excitation spectrum of the normal state
quasiparticles.\cite{schmid,jap} For the case of high temperature
superconductors this behavior is expected to be stronger than in
standard metallic systems due to a quasi-two-dimensional structure
which is responsible for an enhanced critical region around the
superconducting phase transition. This property was used to
explain recent ARPES and tunnelling measurements in cuprates which
experimentally prove the existence of the pseudogap state in their
normal phase.\cite{tremblay,flnonfermi}

\section{Analytical Results in the Non-Fermi Liquid Case}

In this section, we will apply the Ginzburg-Landau functional
formalism to the case of a $s$--wave non--Fermi liquid
superconductor. Our evaluation of normal and anomalous Green's
functions in the superconducting state starts with the hypothesis
that standard Gorkov's equations are still valid. Working on the
same hypothesis Muthukumar et al. \cite{muthu}, following Gorkov's
classic procedure,\cite{gorkov} derived the Ginzburg-Landau
equation and obtained the Ginzburg-Landau coefficients, $A$ and
$B$. Based on these calculations the value of the upper critical
field, $H_c(T)$ near $T_c$ was obtained for the case of the
anomalous Fermi liquid. However, in our analysis we will extend
the calculation to the evaluation of the other Ginzburg-Landau
coefficient, namely $C$, and we will obtain analytical results for
the superconducting state parameters in both spin--charge
separation and anomalous Fermi liquids cases. The role of the mass
renormalization factor, $\b$, is particulary discussed for the
anomalous Fermi liquid case, situation in which a previous
estimation of the Ginzburg-Landau coefficients was made by
Moca.\cite{Moca}

\subsection{The spin-charge separation liquid}

\indent Here, we will focus our attention on the spin--charge
separation liquid case, denoted by the following choice of the
input parameters in Eq. (\ref{nflG}), namely, $\a=0$ and $\eta\neq
1$. The corresponding form of the Green's function is:
\be\label{gscG}
G(\bk,
i\o_n)=\f{1}{(i\o_n-\eta\b\ve_\bk)^{1/2}(i\o_n-\b\ve_\bk)^{1/2}}\;,
\ee
with the mass renormalization factor given by $\b=2/(1+\eta)$. It
is simple to see that for $\eta\ra 1$ the usual form for the Fermi
liquid is recovered.

The critical temperature, $T_c(\eta)$, on this model was first
calculated by Sudbo\cite{sudbo}. A more careful analysis of this
temperature, which also includes the mass renormalization factor,
leads to the following value:
\be\label{tempeta}
T_c(\eta)=\f{2\gamma_E}{\pi}\f{2\o_D}{1+\eta}\exp{
\left[-\f{\pi}{(1+\eta)K(\sqrt{1-\eta^2})}\f{1}{N_0V}\right]}\;,
\ee
where $\gamma_E$ is the Euler constant, $\o_D$ is the Debye
frequency, and $K(k)$ is the complete elliptic integral of the
first kind.

A simple but laborious calculation of the Ginzburg-Landau
parameters leads to the following values:
\be\label{ascs}
A(\eta)=N_0\f{T-T_c(\eta)}{T_c(\eta)}f_A(\eta)\;,
\ee
\be\label{bscs}
B(\eta)=\f{7\zeta(3)N_0}{8\pi^2T^2_c(\eta)}f_B(\eta)\;,
\ee
\be\label{cscs}
C(\eta)=\f{7\zeta(3)N_0v_F^2}{32\pi^2T^2_c(\eta)}f_C(\eta)\;,
\ee
where we introduced the following notations
\be\label{fascs}
f_A(\eta)=\f{(1+\eta)K(\sqrt{1-\eta^2})}{\pi}\;,
\ee
\be\label{fbscs}
f_B(\eta)=\f{1+\eta}{2} F\left(1,\f{1}{2};2;1-\eta^2\right)\;,
\ee
\bea\label{fcscs}
&&f_C(\eta)=\f{1}{2(1+\eta)}\left\{\f{3}{2}\left[\f{3}{2}
F\left(\f{1}{2},\f{1}{2};3;1-\eta^2\right)\right.\right.\nonumber\\
&&+\left.\eta
F\left(\f{3}{2},\f{1}{2};3;1-\eta^2\right)+\f{3\eta^2}{2}
F\left(\f{5}{2},\f{1}{2};3;1-\eta^2\right)\right]\nonumber\\
&&\left.-F\left(\f{1}{2},\f{1}{2};2;1-\eta^2\right)-
\eta^2F\left(\f{3}{2},\f{1}{2};2;1-\eta^2\right)\right\}\;,
\eea
$F(\a,\b;\gamma;z)$ being the hypergeometric function. Making the
limit $\eta\ra 1$ the standard BCS results are recovered.

\indent With all three Ginzburg-Landau coefficients, we can obtain
the corresponding quantities for the superconducting state. The
coherence length can be written as:
\be\label{xiscs}
\f{\xi(\eta,
T)}{\xi_{BCS}(T)}=\f{1}{f_T(\eta)}\sqrt{\f{f_C(\eta)}{f_A(\eta)}}
\sqrt{\f{1-T/T_{c0}}{1-T/[f_T(\eta)T_{c0}]}}\;,
\ee
with
\bea\label{ftscs}
&&f_T(\eta)=\f{T_c(\eta)}{T_{c0}}\nonumber\\&&=\f{2}{1+\eta}
\exp{\left[\left(1-\f{\pi}{(1+\eta)K(\sqrt{1-\eta^2})}\right)\f{1}{N_0V}\right]}\;.
\eea
As we expect $f_T(\eta=1)\ra 1$. In Fig.\ref{figetascs} we plot
the $\eta$-dependence of the ratio between the coherence length in
the spin-charge separation liquid and the standard BCS case. We
observe that, for $\eta \neq 1$, the value of the coherence length
is lower than the one in the standard BCS case. The considered
values of the $T/T_{c0}$ are justified by the range of the
critical region around the transition temperature.

\indent The London penetration depth is obtained as:
\be\label{lscs}
\f{\l(\eta,
T)}{\l_{BCS}(T)}=\sqrt{\f{f_B(\eta)}{f_A(\eta)f_C(\eta)}}
\sqrt{\f{1-T/T_{c0}}{1-T/[f_F(\eta)T_{c0}]}}\;.
\ee
As in the BCS case, the temperature dependence of London
penetration depth and coherence length are the same. In
Fig.\ref{lfigscs}, we plot the $\eta$-dependence of the ratio
between the penetration depth corresponding to the spin-charge
separation liquid and the standard BCS case. As the separation
parameter decreases a lower value for the penetration depth is
obtained.

\indent The value of the specific heat jump at the critical point
for $\eta \neq 1$ was first calculated in Ref. \cite{ioneleur}
based on the Pauli theorem. Using the Ginzburg-Landau
coefficients, one finds:
\be\label{cvscs}
\f{\D C_v(\eta)}{T_c(\eta)}\f{T_{c0}}{\D
C^{BCS}_v}=\f{f_A^2(\eta)}{f_B(\eta)}\;,
\ee
a value which differ from the previous one obtained in Ref.
\cite{ioneleur} by the inclusion of the mass renormalization
factor. In Fig.\ref{cvfigscs}, we plot the $\eta$-dependence of
the specific heat jump at the critical point renormalized by the
same ratio considered in the standard BCS case. One can see that
higher values of the specific heat jump at the transition point
can be expected as the spin--charge separation parameter $\eta$
decreases.

\indent Finally, we are going to evaluate the upper critical
magnetic field, $H_{c2}$. One finds:
\be\label{hscs}
\f{H_{c2}(\eta)}{H^{BCS}_{c2}}=\f{f_A(\eta)
f^2_T(\eta)}{f_C(\eta)} \f{1-T/[f_T(\eta) T_{c0}]}{1-T/T_{c0}}\;,
\ee
leading to a change on the slope of the curve for the upper
critical field near the transition temperature:
\be\label{slopescs}
h(\eta)=\f{\left|\f{dH_{c2}}{dT}\right|_{T=T_c(\eta)}}
{\left|\f{dH_{c2}^{BCS}}{dT}\right|_{T=T_{c0}}}=\f{f_A(\eta)f_T(\eta)}{f_C(\eta)}\;.
\ee
\noindent In Fig.\ref{hfigscs}, we plot the relative value of the
slope of the curve for the upper critical field as function of the
spin-charge separation parameter. An increment of this slope is
predicted as the value of $\eta$ decreases.

\indent At this particular moment we would like to discuss the
presence of a fluctuating vector field, $\vec{a}$, in addition to
the electromagnetic vector field, $\vec{A}$ as it has been
discussed in the literature.\cite{Muthu,Tesa,Weng1,Weng2} We argue
that the full superconducting order parameter, $\D$, calculated in
this paper, for the spin--charge separation case, becomes
\begin{equation}\label{yelling}
\D^{2} \approx \frac{1}{1 + f(\eta)}\left[{\D^{(s)}}^{2} +
f(\eta){\D^{(h)}}^{2}\right],
\end{equation}
\noindent where $f(\eta) \rightarrow 1$ for $\eta \rightarrow 1$
and $\D^{s,h}$ refers to the order parameter for the spinons
(holons), respectively. Along with the assumption of Eq.\
(\ref{yelling}) is implicit that both order parameters are small.
In Eq. (\ref{yelling}), $f(\eta)$ is fixed by the ration between
the two critical temperatures, namely, $T_c^{(s,h)}$. Then, with
these assumptions, we find that our Ginzburg--Landau function can
be expressed as
\bea\label{deltaFs-ch}
F_S-F_N&=&A_s|{\D_\bq}^{(s)}|^2+q^2C_s|{\D_\bq^{(s)}}|^2+\f{B_s}{2}|{\D_\bq^{(s)}}|^4
\nonumber\\ &&+ A_h|{\D_\bq}^{(h)}|^2+q^2C_h|{\D_\bq^{(h)}}|^2
+\f{B_h}{2}|{\D_\bq^{(h)}}|^4\nonumber\\
&&+ B_{s,h}|\Delta^{(s)}|^2 \times |\Delta^{(h)}|^2\;.
\eea
\noindent In Eq. (\ref{deltaFs-ch}), for example, $A_s = A/(1 +
f(\eta))$, etc. Transforming our $G-L$ functional to real space
and substituting the vector potentials, we obtain
\begin{eqnarray}\label{FGL-real}
&&F_S-F_N =\nonumber\\
&& A_s |{\D^{(s)}|}^{2}+C_s|{(-i \nabla - 2\vec{a}^{(s)})
\D^{(s)}}|^2+\f{B_s}{2}|\D^{(s)}|^4\nonumber\\
&& + B_{s,h}|\Delta^{(s)}|^2 \times
|\Delta^{(h)}|^2 \nonumber \\
&&+ A_h|\D^{(h)}|^2 + C_h|{(-i\nabla - \frac{e}{\hbar c}
\vec{A}^{(e)} - \vec{a}^{(h)}) \D^{(h)}}|^2\nonumber\\
&&+\f{B_h}{2}|\D^{(h)}|^4 + \frac{1}{8 \pi} (\nabla \times
\vec{A}^{(e)})^{2} + f_{gauge}\;,
\end{eqnarray}
\noindent where  $f_{gauge}$ is given by
\begin{equation}\label{gauge}
f_{gauge} \equiv \frac{\varrho}{2} (\nabla \times \vec{a})^{2}
~~~~.
\end{equation}
\indent In Eq. (\ref{FGL-real}), $\vec{a}$ is the fluctuating
vector potential (internal gauge field).\cite{Tesa} The factor of
$2$ in front of $\vec{a}$ in the spinon gradient term reflects the
fact that pairs of spinons are assumed to condensate. $f_{gauge}$
describes the dynamics of the internal gauge field, $\vec{a}$. The
internal gauge field, $\vec{a}$, serves only to enforce the local
constraint $b_i^\dagger b_i + f^\dagger_{i,\sigma} f_{i,\sigma} =
1$. The contribution $f_{gauge}$ has been justified by
Sachdev\cite{Sachdev} and Nagaosa--Lee.\cite{Tesa} Franz and
Tesanovich argue that $\varrho$ should be zero in order to
reproduce the experimental data. Eq. (\ref{FGL-real}) is also
similar to Eq. (2) of Franz and Tesanovic.\cite{Tesa} We recover
the results of Eq. (2) of Ref. \cite{Tesa}: we do have a
contribution of the type $\propto {|\D^{(s)}|}^{2} \times
{|\D^{(h)}|}^{2}$, where the coefficient $B_{s,h} = 2B~f_1~f_2$,
where $f_1 = 1/(1+f(\eta))^2$ and $f_2 = f^2(\eta)~f_1^2$.

\indent There is another interpretation due to Muthukumar, Weng
and Sheng.\cite{Muthu} They have two fluctuating gauge fields, one
due to holons and another due to spinons.

\subsection{The anomalous Fermi liquid}

The anomalous Fermi liquid (Eq. (\ref{nflG})) is defined in the
limit $\eta=1$ and $\a\neq 0$, which implies that the normal state
Green's function can be written as:
\be\label{aflG}
G(\bk,
i\o_n)=\f{g(\a)e^{-i\pi\a/2}}{\o_c^\a(i\o_n-\b\ve_\bk)^{1-\a}}\;,
\ee
with the mass renormalization factor given by $\b=1/(1-\a)$. As we
expected, for $\a\ra 0$, the standard Fermi liquid theory is
recovered.

A superconducting phase transition in this system occurs at a
critical temperature:\cite{massren}
\bea\label{tcafl}
T_c^{2\a}(\a)&=&\f{1}{M(\a)}\left[N(\a)\left(\f{\o_D}{1-\a}\right)^{2\a}\right.\nonumber\\
&&\left.-\f{1}{(1-\a)g^2(\a)}\f{\o_c^{2\a}}{P(\a)N_0V}\right]\;,
\eea
only if the value of the attractive interaction  is higher than a
certain critical value,\cite{massren} $V>V_{cr}$. The constants
entering Eq. (\ref{tcafl}) are given by
$P(\a)=2^{2\a}\sin{[\pi(1-\a)]}/\pi$,
$M(\a)=\Gamma^2(\a)[1-2^{1-2\a}]\zeta(2\a)$, and
$N(\a)=\Gamma(1-2\a)\Gamma(\a)/[2\a\Gamma(1-\a)]$, $\Gamma(x)$
being the gamma function. Despite the fact that the value of the
critical temperature is much more complicated in this case, the
standard BCS value still can be obtained as $\a\ra
0$.\cite{IonelPh,nelu}

\indent The first attempt to calculate the Ginzburg--Landau
coefficients for the case of the anomalous Fermi liquid was made
by Moca,\cite{Moca} but our analysis is justified by the necessity
of the mass renormalization factor (omitted in Ref. \cite{Moca})
and by some misprints in the reported results. However, we also
estimate the value of the magnetic upper critical field, $H_{c2}$.

\indent Following the same procedure applied previously to the
spin-charge separation liquid, one finds that the Ginzburg--Landau
parameters are expressed as:
\be\label{aafl}
A(\a)=N_0\f{T-T_c(\a)}{T_c(\a)}f_A(\a)\;,
\ee
\be\label{bafl}
B(\a)=\f{7\zeta(3)N_0}{8\pi^2T^2_c(\a)}f_B(\a)\;,
\ee
\be\label{cafl}
C(\a)=\f{7\zeta(3)N_0v_F^2}{32\pi^2T^2_c(\a)}f_C(\a)\;,
\ee
where we introduced the following functions:
\bea\label{faafl}
&&f_A(\a)=\nonumber\\
&&2\a(1-\a)g^2(\a)P(\a)M(\a)\left[\f{N(\a)}{M(\a)}\right]
\left[\f{\o_D}{(1-\a)\o_c}\right]^{2\a}\nonumber\\
&&\times\left[1-\f{1}{N_0V}\f{g^{-2}(\a)}{(1-\a)P(\a)N(\a)}
\left(\f{(1-\a)\o_c}{\o_D}\right)^{2\a}\right]\;,\nonumber\\
\eea
\bea\label{fbafl}
&&f_B(\a)=\f{2(1-\a)B\left(\f{1}{2},
\f{3}{2}-\a\right)}{\pi}\f{2^{3-4\a}-1}{7}\f{\zeta(3-4\a)}{\zeta(3)}\nonumber\\
&&\times g^4(\a)\left(\f{2\pi\o_D}{(1-\a)\o_c}\right)^{4\a}
\left[\f{N(\a)}{M(\a)}\right]^2\nonumber\\
&&\times \left[1-\f{1}{N_0V}\f{g^{-2}(\a)}{(1-\a)P(\a)N(\a)}
\left(\f{(1-\a)\o_c}{\o_D}\right)^{2\a}\right]^2\;,
\eea
\bea\label{fcafl}
&&f_C(\a)=\f{g^2(\a)\cos{[\pi\a]}}{1-\a}\nonumber\\
&&\times\f{2^{3-2\a}-1}{7}\f{\zeta(3-2\a)}{\zeta(3)}
\f{N(\a)}{M(\a)}\left[\f{2\pi\o_D}{(1-\a)\o_c}\right]^{2\a}\nonumber\\
&&\times \f{2(1-\a)(2-\a)B\left(\f{1}{2},
\f{5}{2}-\a\right)-(1-\a)B\left(\f{1}{2},
\f{3}{2}-\a\right)}{\pi}\nonumber\\
&&\times\left[1-\f{1}{N_0V}\f{g^{-2}(\a)}{(1-\a)P(\a)N(\a)}
\left(\f{(1-\a)\o_c}{\o_D}\right)^{2\a}\right]\;,
\eea
where $B(x, y)$ represents the beta function. Note that the first
two coefficients, despite the mass renormalization factor, are the
same as the ones reported in Ref. \cite{Moca}, whereas the last
one differs from the one already reported. The superconducting
state properties can be evaluated by introducing a new function,
$f_T(\a)$, defined as the ratio of the critical temperature in the
anomalous Fermi liquid and the one corresponding to the standard
BCS case:
\bea\label{ftafl}
&&f_T(\a)=\f{\pi}{2\gamma_E(1-\a)}\left[\f{N(\a)}{M(\a)}\right]^{1/2\a}\times\nonumber\\
&&\exp{\left\{\left[1-\f{1}{N_0V}\f{g^{-2}(\a)}{(1-\a)P(\a)N(\a)}
\left(\f{(1-\a)\o_c}{\o_D}\right)^{2\a}\right]^{1/2}\right\}}\;.\nonumber\\
\eea
In this expression we made use of the exponential form of the
critical temperature reported by Grosu et al.\cite{nelu} To
express the superconducting state properties one can make used of
the previous expression obtained for the spin-charge separation
liquid with the specification that all the functions $f_i(\eta)$
should be replaced by their correspondent in the anomalous Fermi
liquid $f_i(\a)$ ($i=A,B,C,T$).

\indent In Fig.\ref{figxiafl}, we plot the coherence length for
the anomalous Fermi liquid related to the standard BCS value as
function of the non-Fermi parameter $\a$ for different
temperatures in the critical region. The inclusion of the mass
renormalization factor $\gamma=1/(1-a)$ changes completely the
slope of the curve with respect to previous reported
results\cite{Moca}, the coherence length decreasing as $\a$
increases.

\indent Fig.\ref{lfigafl}, shows the ratio of the penetration
depth corresponding to the anomalous Fermi liquid with respect to
the standard BCS value as function of the non--Fermi parameter
$\a$. For most of the interval a decreasing of the penetration
depth as function of the non-Fermi parameter is obtained. The
divergence obtained as $\a$ approaches the limit value $\a\ra 0.5$
could be a simple effect related to the divergence of several
quantities involved in the calculation. A similar effect was
observed also in the critical temperature dependence on the
non-Fermi parameter $\a$.\cite{IonelPh,nelu,muthu}

\indent The dependence of the specific heat jump at the transition
temperature is plot in Fig.\ref{cvfigafl}. Our results show that
smaller values than in the standard BCS case can be achieved for
$0< \a \leq 0.3$ . This conclusion was also obtained in Ref.
\cite{Moca}. However, some changes due to the inclusion of the
mass renormalization factor can be seen in our plot.

\indent A first discussion on the magnetic upper critical field
was made by Muthukumar et al.\cite{muthu} as function of the
non-Fermi parameter $\a$. However, due to the incorrect form of
the Green's function used in their calculation a qualitatively
different dependence of the upper critical field as function of
the non-Fermi parameter $\a$ is expected. In Fig.\ref{hfigafl}, we
plot the slope of the magnetic upper critical field as function of
the non--Fermi parameter $\a$, showing that an increment of this
parameter occurs as $\a$ increases.

\section{Conclusions and discussions}

\indent In this paper, we have discussed the superconducting state
properties in systems described by a non-Fermi liquid normal
state. We specifically analyzed two limit cases corresponding to
the spin-charge separation and anomalous Fermi liquid cases. Our
analysis is based on the evaluation of the Ginzburg-Landau
coefficients corresponding to the superconducting phase
transition, which are used to evaluate the coherence length, the
London penetration depth, the specific heat jump and the magnetic
upper critical field. By comparing our results with experimental
data on high temperature superconductors we could fix the two
parameters, $\eta$ and $\a$, entering the model. For the case of
high temperature superconductors, the experimental data should be
considered in the underdoped region of the phase diagram, where is
well known that the non-Fermi character of the system is much
stronger than in the overdoped regime.

\begin{figure}[tbh]
\centering \scalebox{0.30}[0.30]{\includegraphics*{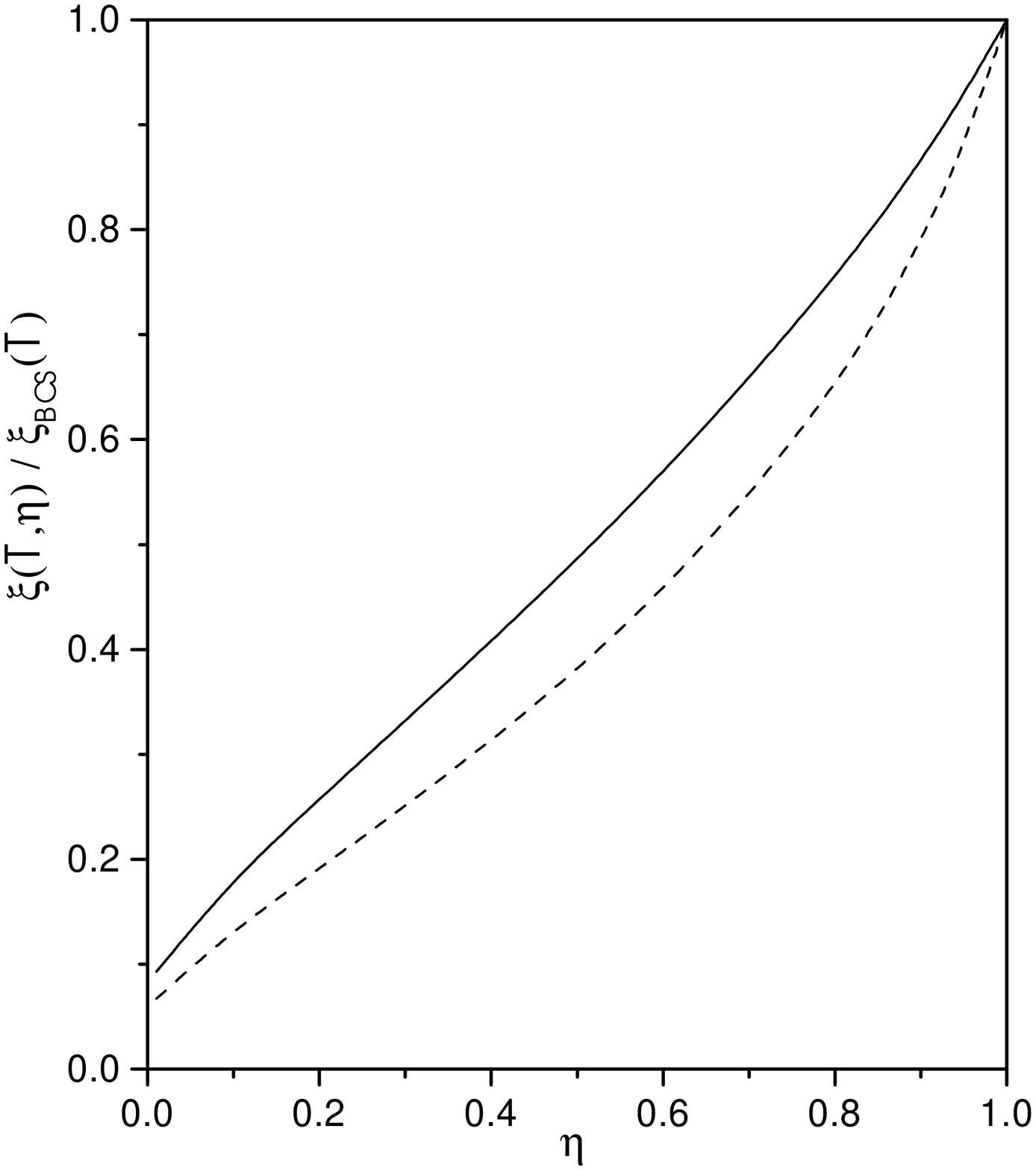}}
\caption{The coherence length ratio as function of the spin-charge
separation parameter $\eta$ for different values of the $T/T_{c0}$
ratio. The full line correspond to a value $T/T_{c0}=0.8$, and the
dashed line to $T/T_{c0}=0.9$.} \label{figetascs}
\end{figure}

\begin{figure}[tbh]
\centering \scalebox{0.30}[0.30]{\includegraphics*{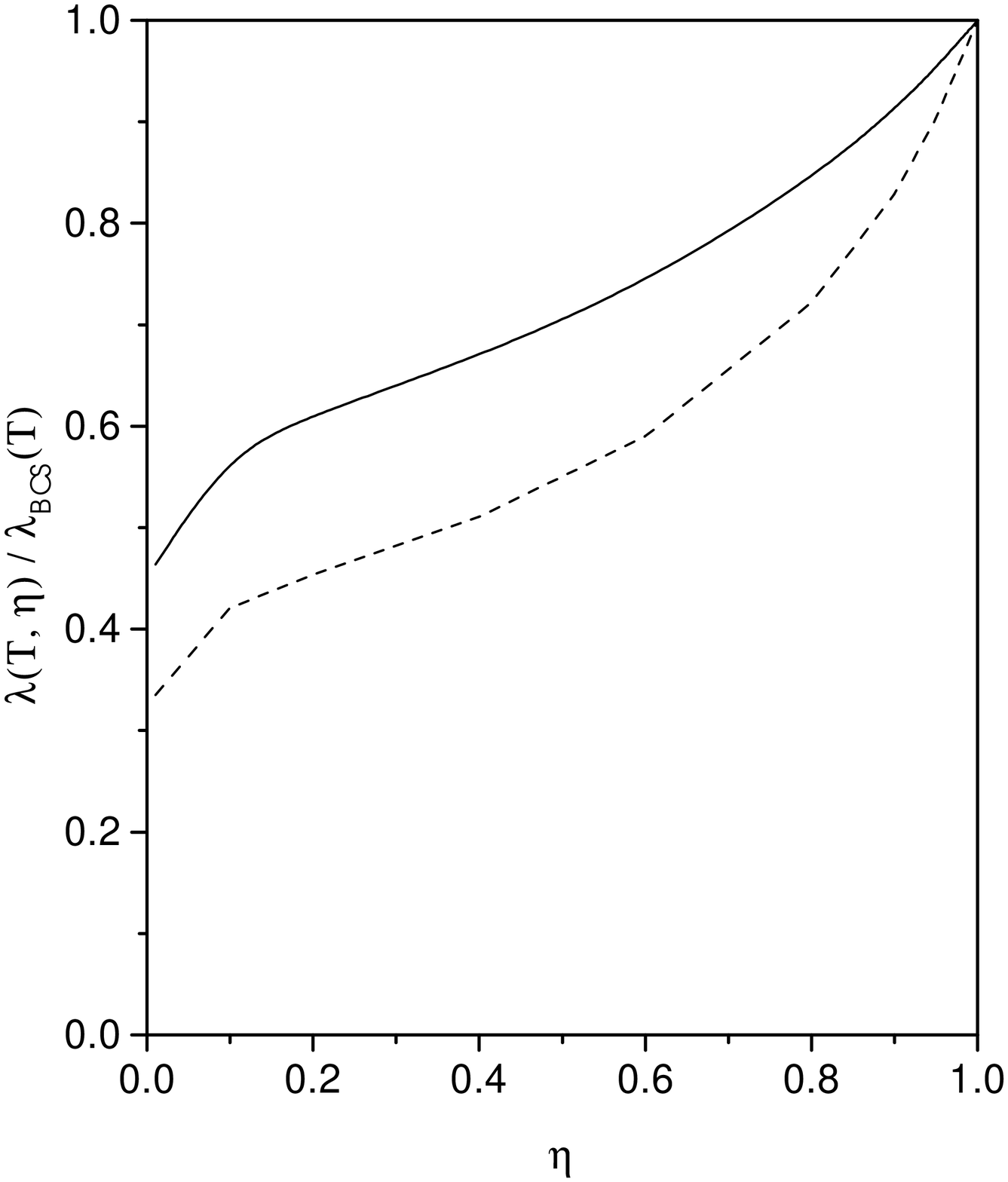}}
\caption{The penetration depth ratio as function of the
spin-charge separation parameter $\eta$ for different values of
the $T/T_{c0}$ ratio. The full line correspond to a value
$T/T_{c0}=0.8$, and the dashed line to $T/T_{c0}=0.9$.}
\label{lfigscs}
\end{figure}

\begin{figure}[tbh]
\centering \scalebox{0.30}[0.30]{\includegraphics*{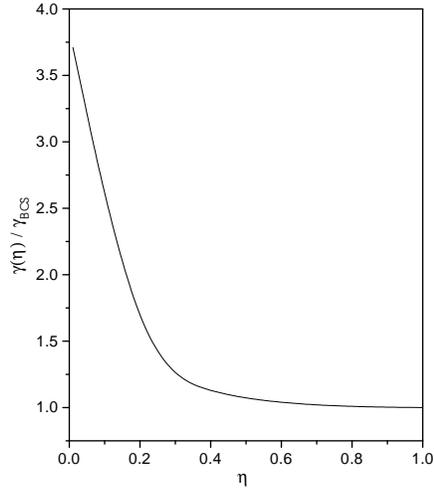}}
\caption{The specific heat jump at the critical point ratio as
function of the spin-charge separation parameter $\eta$.}
\label{cvfigscs}
\end{figure}

\begin{figure}[tbh]
\centering \scalebox{0.30}[0.30]{\includegraphics*{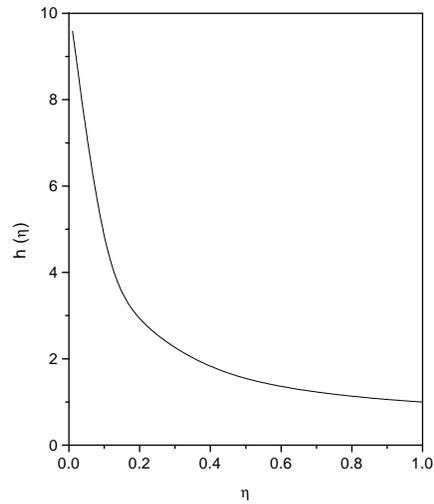}}
\caption{The relative slope of the curve for the magnetic upper
critical field as function of the spin-charge separation parameter
$\eta$.} \label{hfigscs}
\end{figure}

\begin{figure}[tbh]
\centering \scalebox{0.30}[0.30]{\includegraphics*{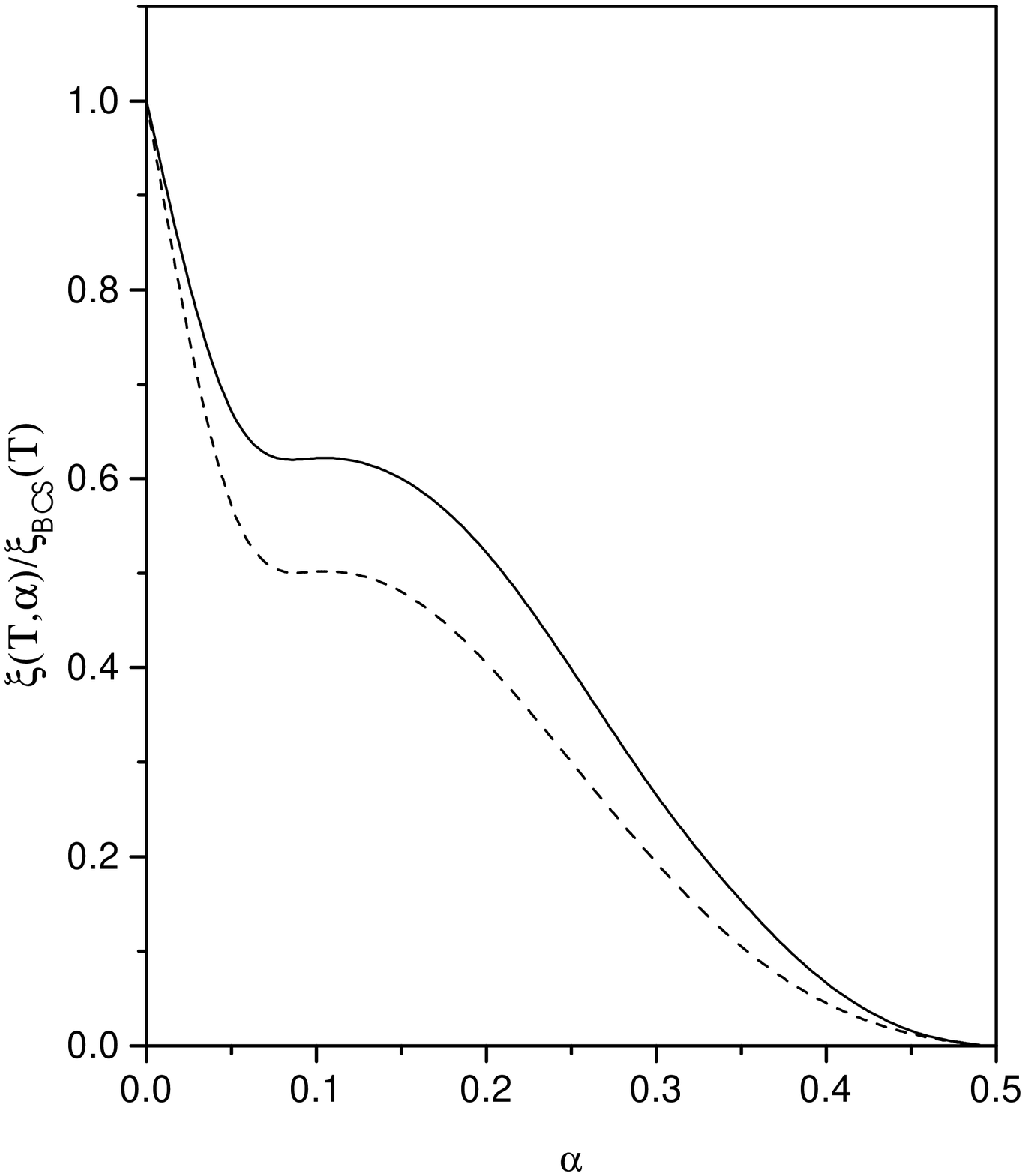}}
\caption{The coherence length ratio as function of the non-Fermi
parameter $\a$ for different values of the $T/T_{c0}$ ratio. The
full line correspond to a value $T/T_{c0}=0.8$, and the dashed
line to $T/T_{c0}=0.9$.} \label{figxiafl}
\end{figure}

\begin{figure}[tbh]
\centering \scalebox{0.30}[0.30]{\includegraphics*{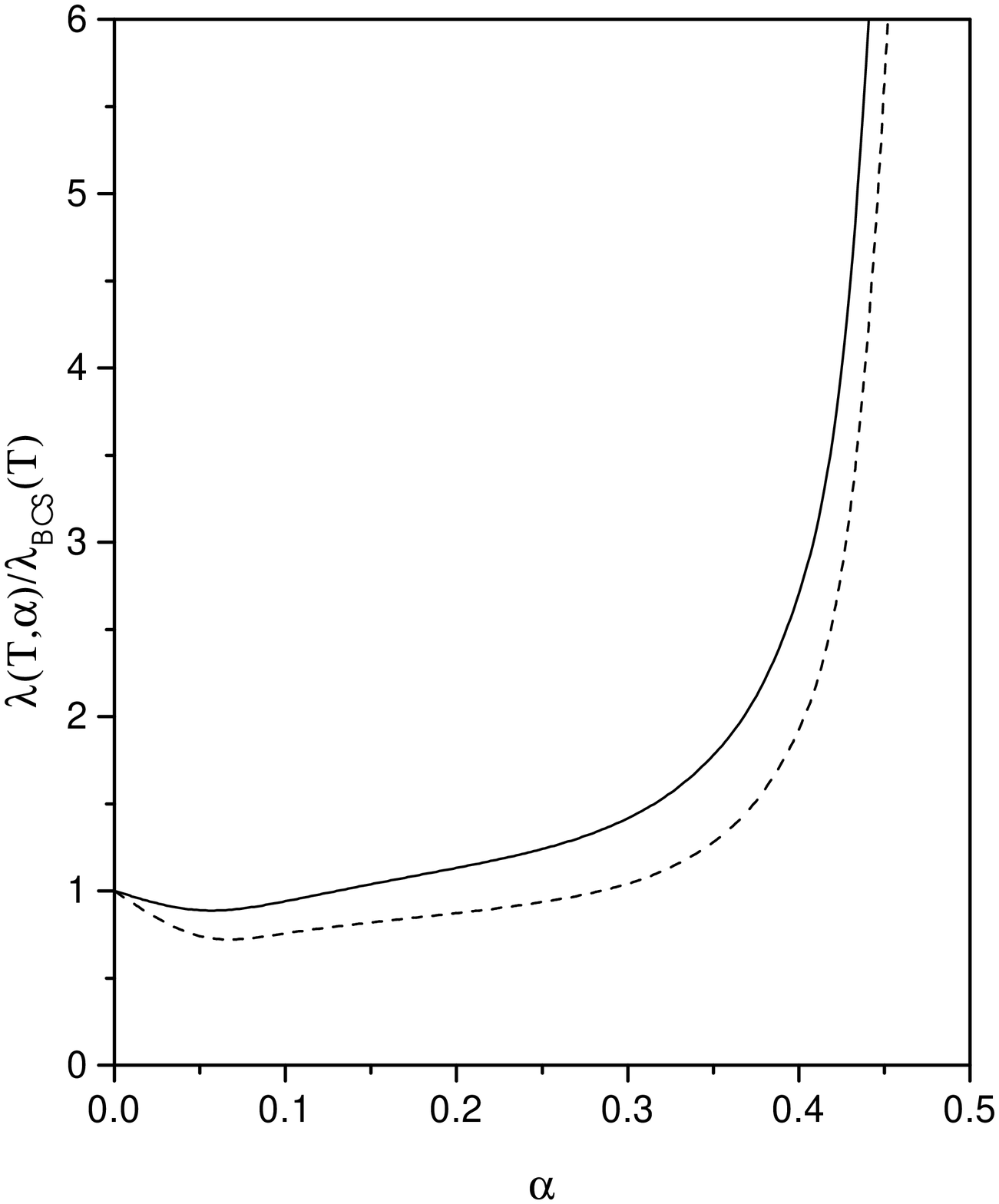}}
\caption{The penetration depth ratio as function of the non-Fermi
parameter $\a$ for different values of the $T/T_{c0}$ ratio. The
full line correspond to a value $T/T_{c0}=0.8$, and the dashed
line to $T/T_{c0}=0.9$.} \label{lfigafl}
\end{figure}

\begin{figure}[tbh]
\centering \scalebox{0.30}[0.30]{\includegraphics*{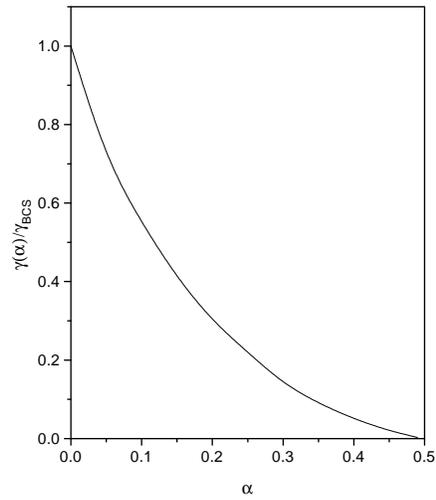}}
\caption{The specific heat jump at the critical point ratio as
function of the non-Fermi parameter $\a$.} \label{cvfigafl}
\end{figure}

\begin{figure}[tbh]
\centering \scalebox{0.30}[0.30]{\includegraphics*{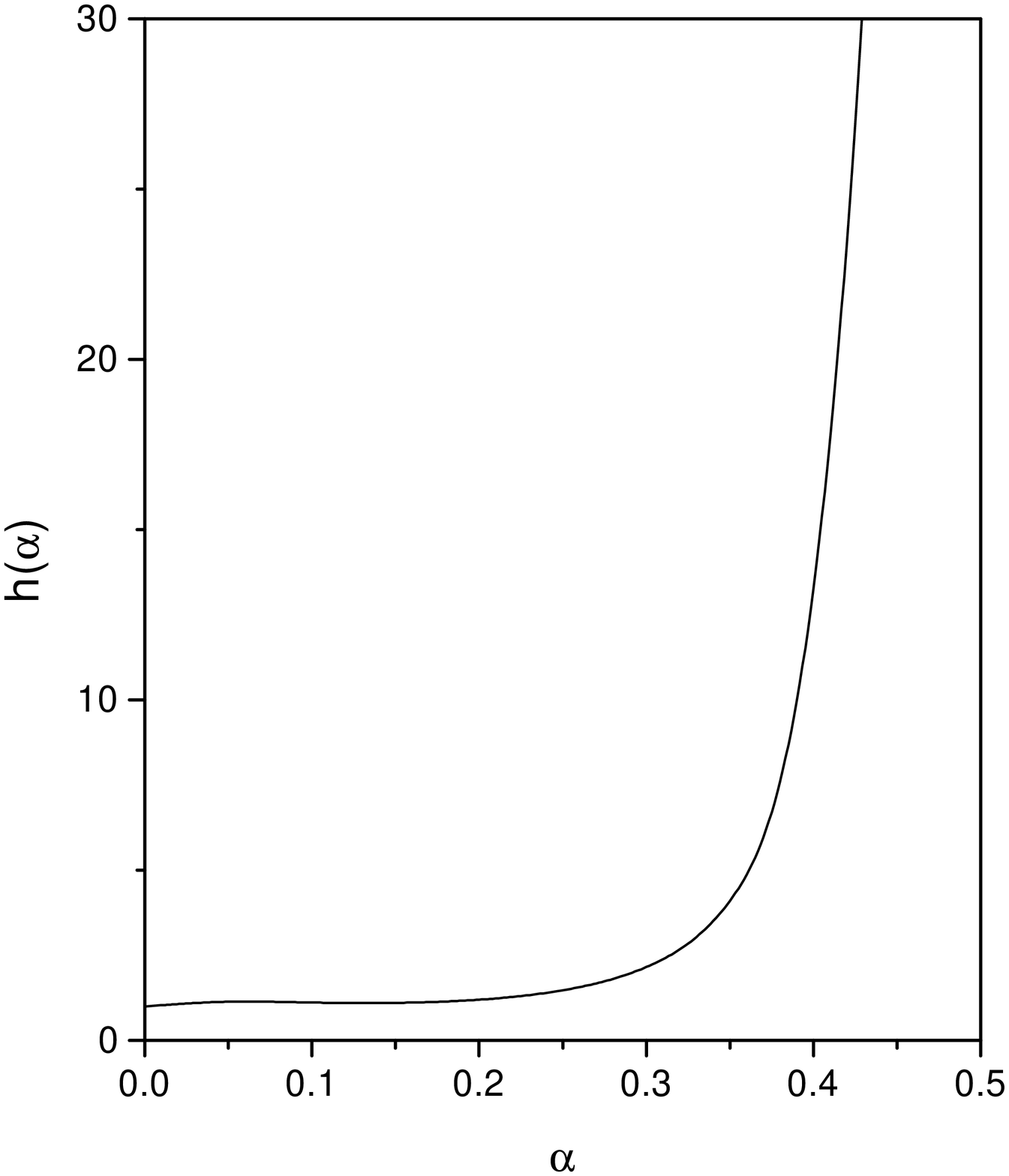}}
\caption{The relative slope of the curve for the magnetic upper
critical field as function of the non-Fermi parameter $\a$.}
\label{hfigafl}
\end{figure}

\indent In general, some similarities can be observed in both
limits of the non--Fermi liquid description. Let us turn our
attention on the coherence length. It is well known that in high
temperature superconductors, the value of this parameter is much
smaller than the one corresponding to standard BCS
superconductors.\cite{coherence} We can see from
Figs.\ref{figetascs} and \ref{figxiafl} that such a behavior can
be achieved when the spin--charge separation parameter $\eta$
decreases, or as the non-Fermi parameter $\a$ increases. Basically
both situations have to do with a stronger non-Fermi character of
the system. The fact that the coherence length decreases to small
values can be interpreted also in terms of a crossover problem,
where in place of varying the attractive interaction leading to
the formation of the Cooper pair, one can vary the non-Fermi
parameter of the problem with the same result, which clearly
identify the fact that the non-Fermi character of the systems is
done by the interaction between the component particles.

\indent In both cases, the standard Fermi liquid can be obtained
by setting $\eta=1$ or $\a=0$. For the anomalous Fermi liquid case
a direct comparison of the specific heat data with experimental
results for high temperature superconductors\cite{loram} shows
that $\alpha$ should satisfy the condition $0.2<\a<0.4$. This is
because the specific heat jump at the critical temperature has
smaller values than in the BCS case. Such values for the non-Fermi
parameter $\alpha$ agree with the observed experimental data for
the penetration depth.\cite{pana} One can see in this case that a
strange divergence of different physical quantities occur once the
value $\alpha\ra 0.5$ is approached. This unphysical result is a
consequence of the various mathematical approximations which we
used on the calculation of these parameters. However, direct
comparisons with the experimental data exclude this value as a
possible correct value for the non-Fermi parameter $\a$. A similar
analysis can be done also for the spin-charge separation liquid.

\indent In short, we have to note that despite the good agreement
between our theoretical results and the experimental data one
cannot conclude that an extension of the Luttinger liquid theory
in two dimensions is the correct answer for the high temperature
superconductivity, at least as long as a direct microscopic theory
can not prove the validity of the phenomenological Green's
function used in the model.

\begin{acknowledgments}
We are very grateful to V.\ Muthukumar, A.\ A.\ Schmidt, L.\
Craco, E.\ V.\ L.\ de Mello, H.\ Beck, and S.\ G.\ Magalh\~aes for
interesting discussions. We thank $C.D.C.H.$--$U.C.$--Venezuela,
$FONACYT$--Venezuela and $FAPERG$--Brasil for finantial support.
One of the authors ($J.J.R.N.$) is a Fellow of the Venezuelan
Program of Scientific Research and a Visiting Scientist at
$IVIC$--Venezuela. We thank M.\ D.\ Garc\'{\i}a--Gonz\'alez for
helping us with the preparation of this manuscript.
\end{acknowledgments}


\begin{thebibliography}{99}
\bibitem{BM}G. Bednorz and K.A. M\"uller, Z.\ Phys. B {\bf 64}, 189 (1986).
\bibitem{theory} See, for example, P. Brusov, {\it Mechanisms of High
Temperature Superconductivity} (Rostov State University
Publishing, 1999).
\bibitem{TesaQED3} M. Franz and Z. Tesanovic,
Phys.\ Rev.\ Lett. {\bf 87}, 257003 (2001).
\bibitem{PWA}P.W. Anderson, {\it The Theory of Superconductivity in
High--$T_c$ Cuprates}. (Princeton, 1997).
\bibitem{Anderson}P.W. Anderson, Science {\bf 256}, 1526 (1992).
\bibitem{yin} L. Yin and S. Chakravarty, Int. J. Mod. Phys. B {\bf
10}, 805 (1996).
\bibitem{massren} J.J. Rodr\'{\i}guez--N\'u\~nez, I. \c{T}ifrea, and
S.G. Magalh\~aes, Phys. Rev. B {\bf 62}, 4026 (2000).
\bibitem{Nolting} W.\ Nolting, Z. Phys. {\bf 255}, 25 (1972); H. Herrmann
and W. Nolting, J. Magn. Magn. Mat. {\bf 170}, 253 (1997).
\bibitem{IonelPh} I. \c{T}ifrea, Ph. Thesis, University of Cluj, 1998
(unpublished).
\bibitem{nelu} I. Grosu, I. \c{T}ifrea, M. Crisan, and S. Yoksan, Phys. Rev.
B {\bf 56}, 8298 (1997).
\bibitem{muthu} V.N. Muthukumar, D. Sa and M.
Sardar, Phys. Rev. B {\bf 52}, 9647 (1995).
\bibitem{sudbo} A.
Sudbo, Phys. Rev. Lett. {\bf 74}, 2575 (1995).
\bibitem{flnonfermi} M. Crisan, C.P. Moca, and I. \c{T}ifrea, Phys.
Rev. B {\bf 59}, 14680 (1999).
\bibitem{Moca} C.P. Moca, Physica C {\bf 349}, 113 (2001).
\bibitem{sadovskii} A.I. Posazhennikova and M.V. Sadovskii, Zh.
Eksp. Theor. Fiz {\bf 115}, 632 (1999) [JETP {\bf 88}, 347
(1999)].
\bibitem{gorkov} See for example, A.L. Fetter and J.D. Walecka,
{\em Quantum Theory of Many-Particle Systems} (McGraw-Hill, 1971).
\bibitem{tsuei} C.C. Tsuei and J.R. Kirtley, Rev. Mod. Phys. {\bf
72}, 969 (2000).
\bibitem{schmid} A. Schmid, Z. Physik {\bf 231}, 324 (1970).
\bibitem{jap} S. Takada, Prog. Theor. Phys. {\bf 44}, 298 (1970).
\bibitem{tremblay} Y.M. Vilk and A.-M.S. Tremblay, J. Phys. I {\bf
7}, 1309 (1997).
\bibitem{ioneleur} I. \c{T}ifrea and M. Crisan, Eur. Phys. J. B {\bf
4}, 175 (1998).
\bibitem{Muthu} V. N. Muthukumar, Z. Y. Weng and D. N. Sheng,
cond-mat/0106225; ibidem, cond--mat/0112339; Z.\ Y.\ Weng and V.\
N.\ Muthukumar, cond--mat/0202079.
\bibitem{Tesa} M. Franz and Z. Tesanovic,
Phys. Rev. B {\bf 63}, 064516 (2001); P. A. Lee and N. Nagaosa,
Phys. Rev. B {\bf 46}, 5621 (1992).
\bibitem{Weng1} Z.\ Y.\ Weng, D. N. Sheng, and C. S. Ting, Phys.
Rev. B {\bf 61}, 12328 (2000-II); ibidem, Phys. Rev. Lett. {\bf
80}, 5401 (1998).
\bibitem{Weng2} Z.\ Y.\ Weng, V. N. Muthukumar, D. N. Sheng, and C. S. Ting,
Phys. Rev. B {\bf 63}, 075102 (2001).
\bibitem{Sachdev}
S. Sachdev, Phys.\ Rev.\ B {\bf 45}, 389 (1992).
\bibitem{coherence} T.M. Riseman, J.H. Brewer, K.H. Chow, W.N. Hardy,
R.F. Kiefl, S.R. Kreitzman, R. Liang, W.A. MacFarlane, P. Mendels,
G.D. Morris, J. Rammer, J.W. Schneider, C. Niedermayer, and S.L.
Lee, Phys. Rev. B {\bf 52}, 10569 (1995).
\bibitem{loram} J.L. Tallon and J.W. Loram, Physica C {\bf 349}, 53
(2001).
\bibitem{pana} C. Panagopoulos, J.R. Cooper, and T. Xiang, Phys.
Rev. B {\bf 57}, 13 422 (1998).
\end{thebibliography}
\end{document}